\documentclass{article}
\usepackage{arxiv}

\usepackage[utf8]{inputenc} % allow utf-8 input
\usepackage[T1]{fontenc}    % use 8-bit T1 fonts
\usepackage{hyperref}       % hyperlinks
\usepackage{url}            % simple URL typesetting
\usepackage{booktabs}       % professional-quality tables
\usepackage{amsfonts}       % blackboard math symbols
\usepackage{nicefrac}       % compact symbols for 1/2, etc.
\usepackage{microtype}      % microtypography
\usepackage{lipsum}
\usepackage[utf8]{inputenc}
\usepackage{subfigure}
\usepackage {caption} 
\usepackage{amsthm}
\usepackage{blindtext}
\usepackage[underline=false]{pgf-umlsd}
\newcommand{\bloodymess}[7][0]{
  \stepcounter{seqlevel}
  \path
  (#2)+(0,-\theseqlevel*\unitfactor-0.7*\unitfactor) node (mess from) {};
  \addtocounter{seqlevel}{#1}
  \path
  (#4)+(0,-\theseqlevel*\unitfactor-0.7*\unitfactor) node (mess to) {};
  \draw[->,>=angle 60] (mess from) -- (mess to) node[midway, above]
  {#3};

  \if R#5
    \node (#3 from) at (mess from) {\llap{#6~}};
    \node (#3 to) at (mess to) {\rlap{~#7}};
  \else\if L#5
         \node (#3 from) at (mess from) {\rlap{~#6}};
         \node (#3 to) at (mess to) {\llap{#7~}};
       \else
         \node (#3 from) at (mess from) {#6};
         \node (#3 to) at (mess to) {#7};
       \fi
  \fi
}
\usepackage{enumerate}
\usepackage{cite}
\usepackage{graphicx}
\usepackage{multirow}
\usepackage{comment}
\usepackage{mathtools}
\usepackage{lmodern}
\usepackage{verbatim} 
\usepackage{float}
\usepackage{tabularx}
\usepackage{amsmath,amssymb}
\usepackage{geometry}
\usepackage{listings}
\usepackage{color}
\usepackage{rotating}
\usepackage{listings}
\usepackage{mathtools}

\usepackage{array}
\newcolumntype{L}[1]{>{\raggedright\let\newline\\\arraybackslash\hspace{0pt}}m{#1}}
\newcolumntype{C}[1]{>{\centering\let\newline\\\arraybackslash\hspace{0pt}}m{#1}}
\newcolumntype{R}[1]{>{\raggedleft\let\newline\\\arraybackslash\hspace{0pt}}m{#1}}
\usepackage{geometry}
\usepackage{booktabs}

\title{Non-invertible Anonymous Communication for the Quantum Era}

\author{
  Luis A.~Lizama\thanks{https://www.upp.edu.mx/posgrado---%\emph{not} for acknowledging funding agencies.
  } \\
  Dirección de Investigación, Innovación y Posgrado\\
  Universidad Politécnica de Pachuca\\
  Hgo., México\\
  \texttt{luislizama@upp.edu.mx} \\
}

\begin{document}
\maketitle

\begin{abstract}
We introduce a new approach for circuit anonymous communication based on Lizama's non-invertible Key Exchange Protocol (ni-KEP) which has been conceived to work in the quantum era. Lizama's protocol has the smallest key size when compared to main post-quantum schemes thus it becomes a promising alternative for the quantum era. Circuit-based communication can be scaled to support the Hidden Service Protocol (HSP) as well as cross-domain digital certificates that promise greater computing security, speed and efficiency.
\end{abstract}

% keywords can be removed
\keywords{Anonymity \and TOR \and circuit \and non-invertible}

section{Introduction}
\label{sec:Introduction}

The Onion Routing (TOR) protocol provides non-traceability to the users of data networks. Non-traceability also called anonymity, complements other well established data security services as confidentiality, integrity and authentication. On the one hand, non-traceability is imperative for activists and informants, but on the other it does not allow to identify the origin where cybernetic crimes could be committed.

Unfortunately, the security of TOR's underlying algorithms is based on computational problems whose security has been threatened by the imminent development of quantum computers since Peter Shor conceived a useful quantum algorithm to solve in polynomial time the integer factoring problem  running on a hypothetical quantum computer~\cite {shor1994algorithms}. Worse still, the vast majority of the public key cryptography most used today, among which we can mention RSA, Diffie-Hellman (DH) and elliptic curve cryptography (ECC), will become useless in the near future because quantum computers will be able to break it~\cite{barreno2002future}. Because of this, the National Institute of Standards and Technology (NIST) began in 2015 a process of evaluation of post-quantum algorithms to select the most suitable methods for cryptography in the quantum era. Currently, this process is in the third round of evaluation ~\cite{PQC, chen2016report}.

Cryptography in the quantum era can be separated into two main approaches: quantum and post-quantum cryptography. A detailed review of these fields is beyond the purpose of this article. However, let us simply describe that quantum cryptography is based on the principles of quantum physics which are used to establish a secret key between two previously authenticated remote parties~\cite {bennett1984quantum}. In this scheme, the Heisenberg uncertainty principle guarantees that an attacker is unable to control quantum communication because it generates detectable noise. Recently, research results have been published demonstrating the ability to resist quantum attacks~\cite{lizama2014quantum, lizama2016quantum,lizama2020quantum,lizama2021beyond}.

Post-quantum cryptography encompasses those methods designed to be immune to the computational power of quantum computers~\cite{bernstein2017post, chen2016report}. Several algorithms based on computational problems have been invented, the difficulty of which goes beyond the theoretical capacity of quantum computers. Among the most prominent techniques, we can include code-based cryptography~\cite{mceliece1978public,ott2019identifying}, lattice based key exchange~\cite{wang2014lattice,gupta2016cryptanalysis}, supersingular elliptic curve isogeny~\cite{jao2011towards}, hash based cryptography~\cite{lamport1979constructing,merkle1982method,lizama2019digital,Lizama2017public}, zero knowledge~\cite{goldwasser1989knowledge,ben2018scalable} and multivariate cryptography~\cite{matsumoto1988public,ding2005rainbow}.  

As a final remark, we emphasize that the security of current technologies widely used on the internet as bitcoin and blockchain would be seriously threatened by quantum computers. TOR would be in the same disadvantage, whose basic anonymizer mechanism is supported on RSA, Diffie-Hellman and Elliptic Curve Cryptography. In this work, we extend a novel cryptographic approach introduced by Lizama in \cite{lizamanon,lizama2021non} to achieve anonymous communication. In Appendix~\ref{Lizama's KEP} can be found a detailed description of this protocol. The rest of the paper is organized as follows: in Section 2 we describe the fundamentals of TOR and circuit operation, we briefly discuss the hidden service protocol. In Section 3 we explain some security  issues of Lizama's ni-KEP and how it can be used to achieve non-traceability circuit communication. Finally, in Section 4 we emphasize the main advantages of our approach in terms of the key size, scalability and interoperability.

\section{Non-traceable circuit-based communication}

Today, the Onion Routing Protocol (TOR) has become the most widespread technology for achieving non-traceability and anonymous web browsing ~\cite{syverson1999onion}. Non-traceability can be achieved thanks to a fundamental approach known as circuit-based communication between network nodes. In the following paragraphs we will discuss circuit-based communication and the hidden service protocol.

\subsection{Circuit Initialization}

Circuit-based communication between a user Alice that we will denote here as A and a web server that we represent as S, is built along at least three nodes of the network that we write as B, C, D: the input node is denoted here as B, the intermediate node is C and the exit node is represented as D.
The purpose of the circuit is to establish a secret key between Alice and each of the three nodes of the network hiding Alice's identity, at least from the intermediate node and the exit node. Alice establishes the secret key $k_{ab}$ with B, who in turn allows Alice to set $k_{ac}$ with C but as a result of this process input node B cannot know this key. Nodes B and C will allow Alice to derive $k_{ad}$ with D, but again they cannot know the secret key. Using these keys, Alice prepares an onion data packet (see Figure~\Ref{fig:Fig.3}) so that input node B knows the request coming from Alice but ignores the final destination of the packet. Intermediate node C knows the identity of the input node and the next node in the packet path, but neither of them correspond to Alice or the final web server's identity. Finally, exit node D knows where is the web server but is unaware that Alice sent the request.

\begin{figure}[htbp]
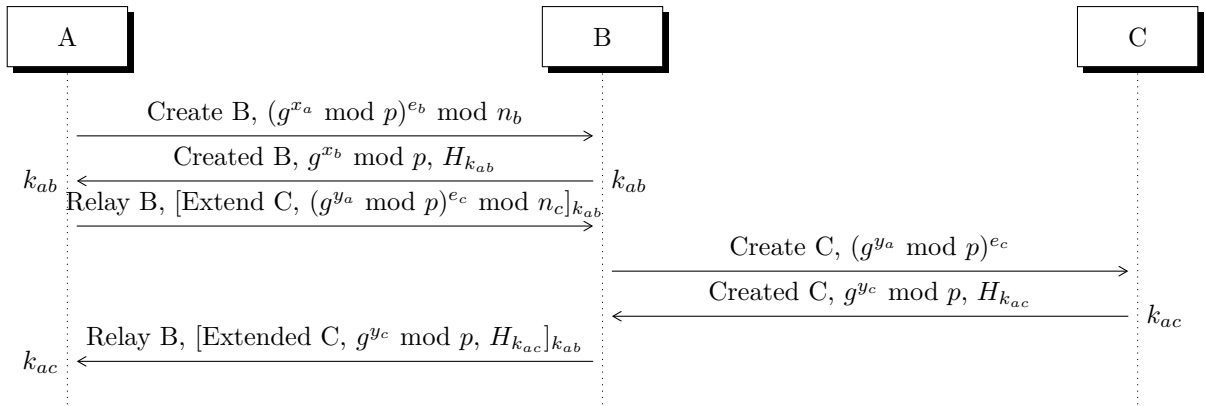
 
\centering
\begin{sequencediagram}
\newinst{A}{A}
\newinst[5.5]{B}{B}
\newinst[5.5]{C}{C}
\bloodymess[0]{A}{Create B, $(g^{x_a}$ mod ${p)^{e_b}}$ mod $n_b$}{B}{R}{}{}
\bloodymess[0]{B}{Created B, $g^{x_b}$ mod $p$, $H_{k_{ab}}$}{A}{L}{$k_{ab}$}{$k_{ab}$}
\bloodymess[0]{A}{Relay B, [Extend C,$ $ $(g^{y_a}$ mod ${p)^{e_c}}$ mod $n_c]_{k_{ab}}$}{B}{R}{}{}
\bloodymess[0]{B}{Create C,$ $ $(g^{y_a}$ mod ${p)^{e_c}}$}{C}{R}{}{}
\bloodymess[0]{C}{Created C, $g^{y_c}$ mod $p$, $H_{k_{ac}}$}{B}{L}{$k_{ac}$}{}
\bloodymess[0]{B}{Relay B, [Extended C, $g^{y_c}$ mod $p$, $H_{k_{ac}}]_{k_{ab}}$}{A}{L}{}{$k_{ac}$}
\end{sequencediagram}
\caption{Alice sends B a packet with a Create B tag which means she must execute the key exchange protocol with A. Then, B sends a Created B confirmation response to Alice. Now, Alice sends B a packet labeled Relay B containing the Extend C command encrypted with $k_{ab}$. B decrypts it and forwards it to C. When B receives the Created C packet from C, she sends it back to Alice as a Relay B packet encrypted with $k_{ab}$. The hash value of the key $H_{k}$ is appended to the packets for Alice to authenticate the source of the keys.}
\label{fig:Fig.1}
\end{figure}

\begin{figure}[htbp]
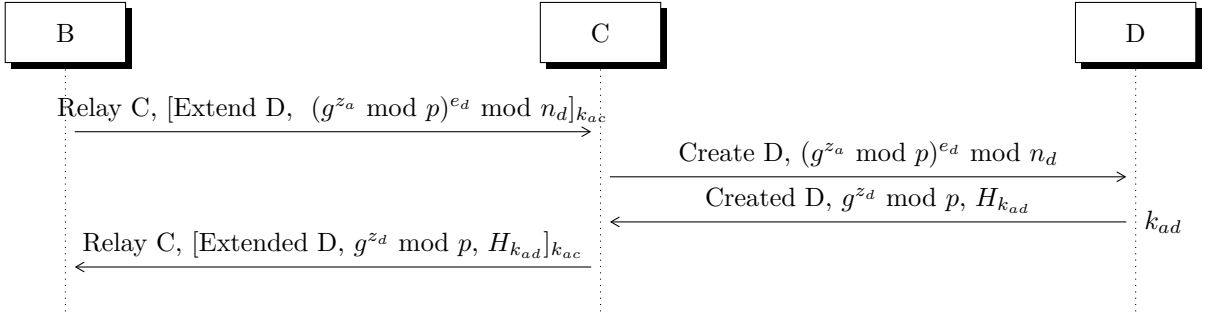
 
\centering
\begin{sequencediagram}
\newinst{B}{B}
\newinst[5.5]{C}{C}
\newinst[5.5]{D}{D}
\bloodymess[0]{B}{Relay C, [Extend D, $ $ $(g^{z_a}$ mod ${p)^{e_d}}$ mod $n_d]_{k_{ac}}$}{C}{R}{}{}
\bloodymess[0]{C}{Create D,$ $ $(g^{z_a}$ mod ${p)^{e_d}}$ mod $n_d$}{D}{R}{}{}
\bloodymess[0]{D}{Created D, $g^{z_d}$ mod $p$, $H_{k_{ad}}$}{C}{L}{$k_{ad}$}{}
\bloodymess[0]{C}{Relay C, [Extended D, $g^{z_d}$ mod $p$, $H_{k_{ad}}]_{k_{ac}}$}{B}{L}{}{}
\end{sequencediagram}
\caption{When B receives a packet labeled Relay B from Alice, it forwards it to C as a Relay C command. Then, C decrypts the message with $k_{ac} $ and forwards it to D as a Create D command. In the opposite direction, when C receives a packet from D labeled Created D, forwards it to B encrypted with $k_{ac} $ and labeled Relay C. Then B delivers it to Alice as a Relay B packet. Alice decrypts it with $k_{ac} $ and she gets $k_{ad}$, the secret key between Alice and output node D.}
\label{fig:Fig.2}
\end{figure}

\begin{figure}[htbp]
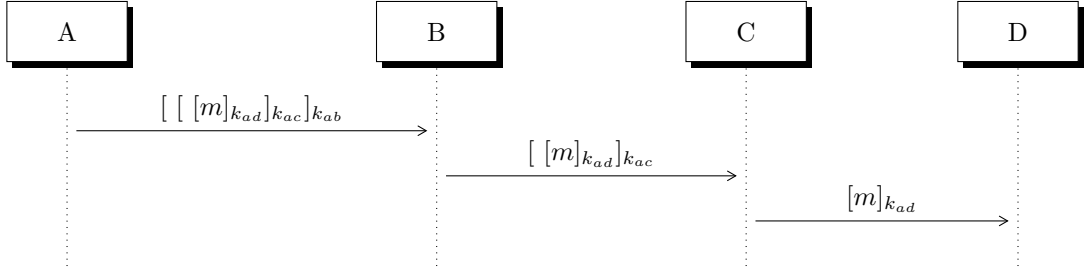
 
\centering
\begin{sequencediagram}
\newinst{A}{A}
\newinst[3.3]{B}{B}
\newinst[2.5]{C}{C}
\newinst[2]{D}{D}
\bloodymess[0]{A}{$[$ $[$ $[ m ]_{k_{ad}} ]_{k_{ac}}]_ {k_{ab}}$}{B}{R}{}{}
\bloodymess[0]{B}{$[$ $[ m ]_{k_{ad}}]_{k_{ac}}$}{C}{R}{}{}
\bloodymess[0]{C}{$[ m ]_{k_{ad}}$}{D}{R}{}{}
\end{sequencediagram}
\caption{Basic functionality of the circuit-based protocol. Node D retrieves $m$ and delivers it to the web server.}
\label{fig:Fig.3}
\end{figure}

When using Diffie-Hellman (DH) as the key exchange algorithm and RSA as the authentication protocol, we see the circuit initialization in Figs.~\ref{fig:Fig.1} and \ref{fig:Fig.2} where $g^{x_i} \ \text{mod} \ p$ represents the DH constructor sent by user $i$ and ${(g^{x_i} \ \text{mod} \ p)}^{e_j} \ \text{mod} n_j$ means that the constructor is encrypted with the public key $(e_j, n_j)$ that belongs to the target user $j$.

In Figure~\Ref{fig:Fig.3} the onion routing protocol on the created circuits is shown. The notation $[m]_k$ denotes that the message $m$ is encrypted with $k$. Therefore, in the final phase of the protocol, the output node D gets $m$ and delivers the request to the server. Then node D returns the server's response to C encrypted with $k_{ad}$ by running the onion routing protocol in reverse.

\subsection{The Hidden Service Protocol}

\begin{figure}[htbp] 
\begin{center}
\includegraphics[height = 5cm]{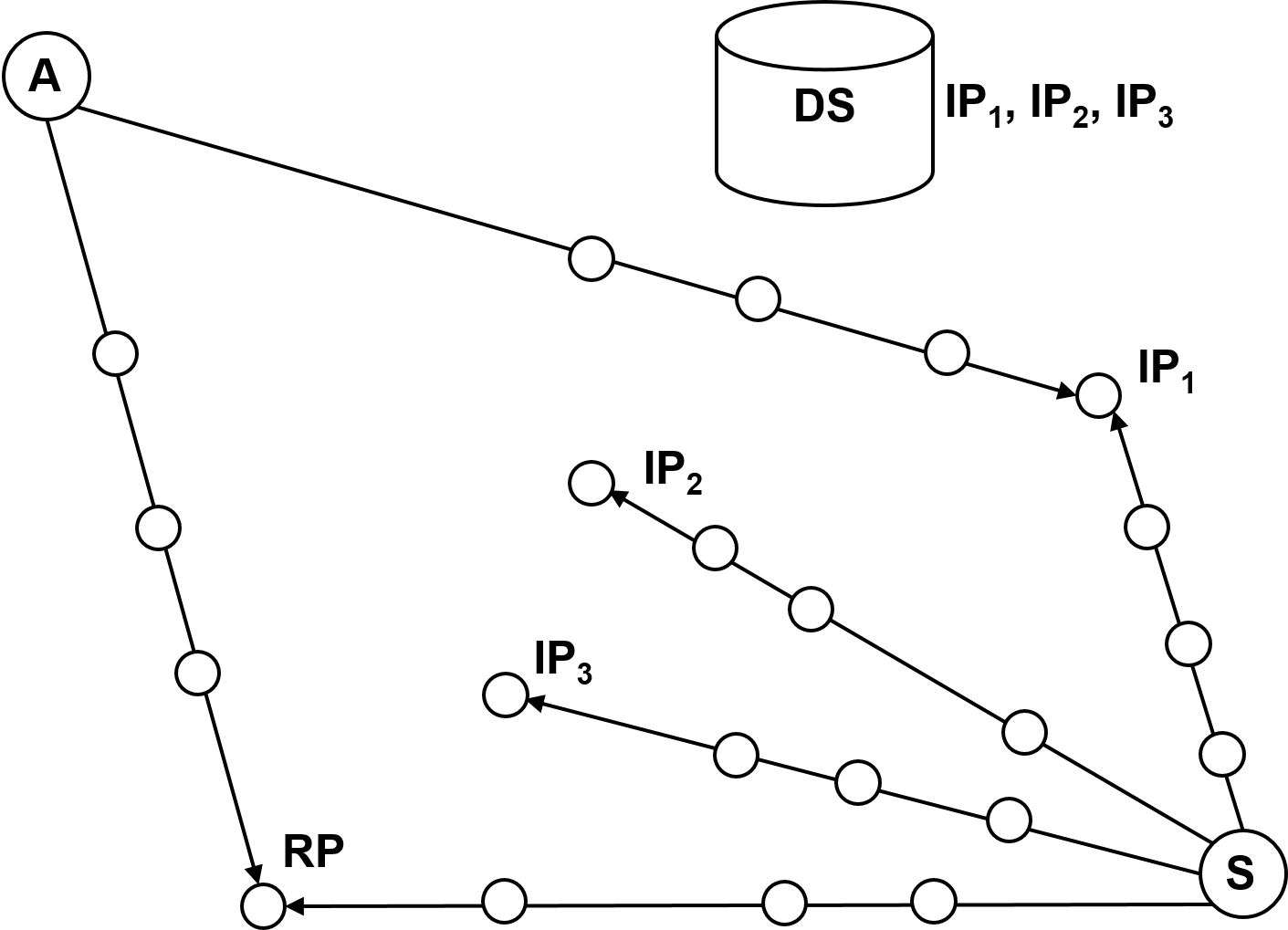}
\caption{Simplified representation of the Hidden Service Protocol (HSP).}
\label{fig:Fig.4}
\end{center}
\end{figure}

In the above discussion, circuit-based communication gives Alice (A) anonymity, however, it is still possible to identify the location of the Web Server (S) on the network. To avoid being traced, the server must run the Hidden Service Protocol (HSP) which is based on basic circuit function and the Onion Routing Protocol. HSP is represented in Figure~\Ref{fig:Fig.4} where the use of the circuits is observed. Instead of revealing its location on the network (that is, the IP address) by means of a directory service (DS), the web server publishes three nodes called Introduction Points (IP) that act as representatives of S and are connected to S through the three usual nodes of circuit-based communication. Alice randomly chooses an IP to inform S the chosen meeting point (RP) where the service will be achieved as shown in Figure~\Ref{fig:Fig.4}. The Hidden Service Protocol demonstrates that the operation of the basic circuit makes it possible to keep the location of users and services on the network anonymous.   

\section{Non-traceability with Lizama's non-invertible Key Exchange Protocol}

A detailed description of Lizama's non-invertible Key Exchange Protocol (ni-KEP) can be found in \cite{lizamanon,lizama2021non}. An overview of the protocol has been written in Appendix~\ref{Lizama's KEP}. In this section we detail a method to construct a circuit using Lizama's algorithm. A previous attempt was presented in~\cite{Lucio2010sistema}, however the protocol is vulnerable to a MITM attack. First of all, let us introduce the mathematical notation we will use to describe the protocol and some important security issues.

\subsection{Mathematical notation}

We use the symbol $(P_i, Q_i)$ to represent the public key of the user $i$. Here, $P_i=p^{2x_i} k_i$ and $Q_i=q^{y_i} k_i$ where $(x_i, k_i)$ constitute the private key of the user $i$ provided that $x_i + y_i = \phi(n) +1$. User $j$ raises the public key of $i$ to the power given by his private key numbers. Then $j$ returns to $i$ the number $[k_{i,j}] \ k_i$ where $[k_{i,j}] = {p^{2x_i x_j}}q^{y_i y_j}$ and $k_i$ is a component of the private key of the user $i$, then he applies the inverse of $k_i$ in order to derive the shared secret key $k_{i,j}$. The same procedure is applied in the opposite direction, so the user $i$ sends to $j$ the number $[k_{i,j}] \ k_j$ to get the same secret number $k_{i,j}$ (see Table~\ref{tab:Table50}).

\begin{table}[htbp] 
\begin{center}
\caption{Mathematical representation of the public keys. All operations are performed module $n$ where $n=pqr$.}
\begin{tabular}{l   c}
\\	
Short notation & \hspace{3mm} Math operation \hspace{3mm} \\
\hline 
$(P_i, Q_i)$  & \hspace{3mm} $P_i=p^{2x_i} k_i$, $Q_i=q^{y_i} k_i$ \\ \\
${P_i}^{x_j} \cdot {Q_i}^{y_j}$  & \hspace{3mm} $\left({p^{2x_i} k_i}\right)^{x_{j}} \cdot \left({q^{y_i} k_i }\right)^{y_j}$ \\ \\
$[k_{i,j}]$ $k_i$ &  \hspace{3mm} ${p^{2x_i x_j}}q^{y_i y_j} k_i$\\ 
\hline
\end{tabular}
\label{tab:Table50}
\end{center}
\end{table}

\subsection{Prefix attack}

Lizama's ni-KEP works as a cryptosystem as described in Appendix~\ref{Lizama's KEP}. Let's just highlight some of the most important security properties:

\begin{itemize}
    \item[1.] Encryption in Lizama's algorithm is performed multiplying by the encryption key, so we denote that a message $m$ is encrypted with the key $k$ as $[w]_{k}$. %Therefore, the multiplication is homomorphic when several keys are applied successively.
    \item [2.] The Lizama cryptosystem is homomorphic under the multiplication operation. Therefore, if we consider the encryption $[w]_{k_a k_b}$ and assuming that we have the appropriate decryption key we can write:
    \begin{itemize}
        \item [] $[w]_{k_a  k_b  {k_a}^{-1}} = [w]_{k_a}$
        \item [] $[w]_{k_a k_b {k_a}^{-1}} = [w]_{k_b}$
    \end{itemize}
\end{itemize}
    
where ${k_i}^{-1}$ represents the multiplicative inverse of $k_i$ in $\mathbb{Z}_{n}$ and $n=pqr$. Special attention must be paid to the messages exchanged through the public channel due to the rules of modular multiplication. When an eavesdropper captures let's say $[w]_{k_m}$ which is a prefix of another that we can write as $[w]_{k_m k_n}$, then she derives $k_n$ because Eve first calculates the inverse of the prefix, it is say ${[w]_{k_m}}^{-1}$ and then she factors it from the second number. However, in Lizama's algorithm, $k_n$ and $[w]_{k_m}$ are non-invertible integers in $\mathbb{Z}_{n}$. Despite this, the attacker could multiply them by ${2}^{-2}$ which changes the modulus from $4r$ to $r$, e.g. $p=q=2$ and $n=4r$. Suppose that Eve has captured from the public channel the integers $[4x]_{k_m}$ and $[4x]_{k_m k_n}$ both in the module $4r$, and then she divides each of them by $2^{-2}$, then Eve gets $k_n$ because $({[x]_{k_m}}) ^{-1} \cdot ({[x]_{k_m k_n}}) = k_n \ \text{mod} \ r$. To avoid a prefix attack, $k_n$ should be chosen to be greater than the prime integer $r$. 

\subsection{Non-traceability with Lizama's ni-KEP}

\begin{figure}[htbp]
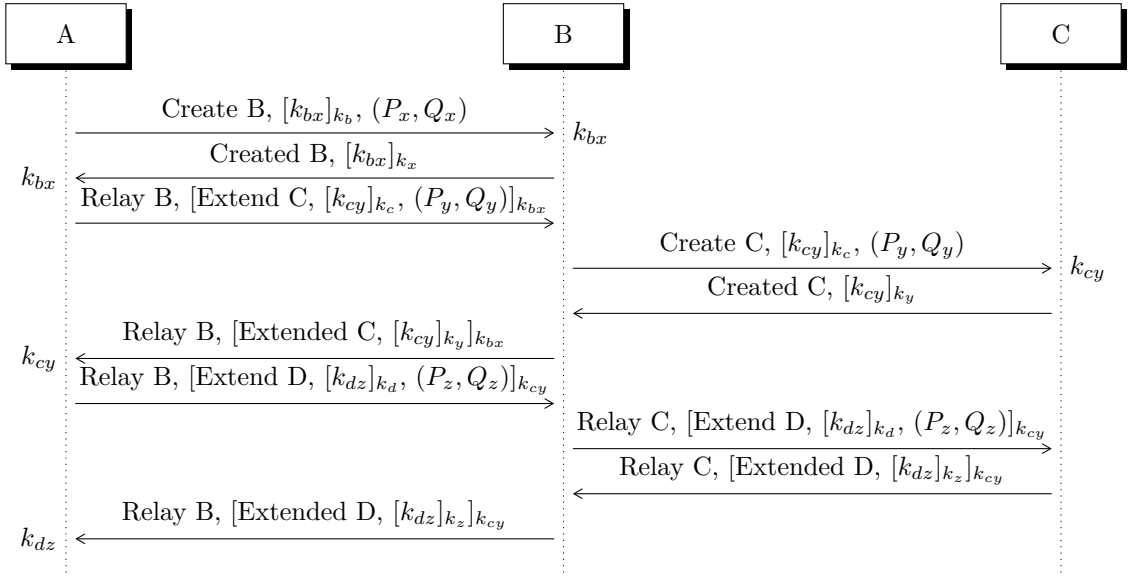
 
\centering
\begin{sequencediagram}
\newinst{A}{A}
\newinst[5]{B}{B}
\newinst[5]{C}{C}
\bloodymess[0]{A}{Create B, $[k_{bx}]_{k_b}$,$ $ ${(P_x, Q_x)}$}{B}{R}{}{$k_{bx}$}
\bloodymess[0]{B}{Created B, $[k_{bx}]_{k_x}$}{A}{L}{}{$k_{bx}$}
\bloodymess[0]{A}{Relay B, [Extend C, $[k_{cy}]_{k_c}$, ${(P_y, Q_y)}]_{k_{bx}}$}{B}{R}{}{}
\bloodymess[0]{B}{Create C, $[k_{cy}]_{k_c}$, ${(P_y, Q_y)}$}{C}{R}{}{$k_{cy}$}
\bloodymess[0]{C}{Created C, $[k_{cy}]_{k_y}$}{B}{R}{}{}
\bloodymess[0]{B}{Relay B, [Extended C, $[k_{cy}]_{k_y}]_{k_{bx}}$}{A}{L}{}{$k_{cy}$}
\bloodymess[0]{A}{Relay B, [Extend D, $[k_{dz}]_{k_d}$, ${(P_z, Q_z)}]_{k_{cy}}$}{B}{R}{}{}
\bloodymess[0]{B}{Relay C, [Extend D, $[k_{dz}]_{k_d}$, ${(P_z, Q_z)}]_{k_{cy}}$}{C}{R}{}{}
\bloodymess[0]{C}{Relay C, [Extended D, $[k_{dz}]_{k_z}]_{k_{cy}}$}{B}{L}{}{}
\bloodymess[0]{B}{Relay B, [Extended D, $[k_{dz}]_{k_z}]_{k_{cy}}$}{A}{L}{}{$k_{dz}$}
\end{sequencediagram}
\caption{Circuit initialization protocol using ni-KEP.}
\label{fig:Fig.5}
\end{figure}

\begin{figure}[htbp]
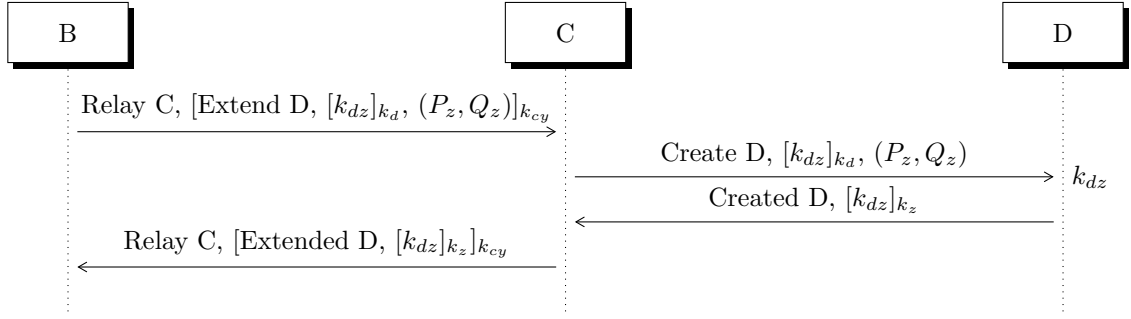
 
\centering
\begin{sequencediagram}
\newinst{B}{B}
\newinst[5]{C}{C}
\newinst[5]{D}{D}
\bloodymess[0]{B}{Relay C, [Extend D, $[k_{dz}]_{k_d}$, ${(P_z, Q_z)}]_{k_{cy}}$}{C}{R}{}{}
\bloodymess[0]{C}{Create D,$ $ $[k_{dz}]_{k_d}$, ${(P_z, Q_z)}$}{D}{R}{}{$k_{dz}$}
\bloodymess[0]{D}{Created D, $[k_{dz}]_{k_z}$}{C}{R}{}{}
\bloodymess[0]{C}{Relay C, [Extended D, $[k_{dz}]_{k_z}]_{k_{cy}}$}{B}{L}{}{}
\end{sequencediagram}
\caption{Circuit initialization (cont).}
\label{fig:Fig.6}
\end{figure}

Lizama's non-traceable protocol is depicted in Figures~\ref{fig:Fig.5} and \ref{fig:Fig.6}. The general idea is that node $i$ can establish a secret key with node $j$ but preserving anonymity of node $i$ because $i$ uses the public constructor $(P_j, Q_j)$ to send $[k_{jw}]_{k_j}$ to $j$ along with the temporary public constructor $(P_w, Q_w)$. Let us describe the steps of the protocol:

\begin{itemize}
    \item [1.] Using B's public key $(P_b, Q_b)$, Alice computes $[k_{bx}]_{k_b}$ and sends it to B along the constructor $(P_x, Q_x)$ labeled as a Create command.
    \item [2.] B derives $k_{bx}$ and he takes $(P_x, Q_x)$ to compute $[k_{bx}]_{k_x}$, then he sends it back to Alice inside a Created command who derives $k_{bx}$, the secret key between A and B. 
    \item [3.] Using C's public key $(P_c, Q_c)$, Alice computes and sends $[k_{cy}]_{k_c}$ to B along the constructor $(P_y, Q_y)$ inside a Relay command and encrypted with $k_{bx}$.
    \item [4.] B decrypts the message with $k_{bx}$ so he gets $[k_{cy}]_{k_c}$, now he knows the identity of the next node and the public constructor $(P_y, Q_y)$, then B forwards them to C inside a Create C command.
    \item [5.] C derives $k_{cy}$ and he responds with $[k_{cy}]_{k_y}$ to B inside a Created C command, then C encrypts the packet with $k_{bx}$ and forwards it to A as a Relay B command. Alice decrypts the command and she derives $k_{cy}$, the secret key between A and C.
    \item [6.] Using D's public key $(P_d, Q_d)$, Alice computes $[k_{dz}]_{k_d}$ and sends them to B along the constructor $(P_z, Q_z)$ encrypted with $k_{cy}$ inside a Relay B command. B forwards the packet to C as a Relay C command, then C decrypts the message using $k_{cy}$.   
    \item [7.] C gets $[k_{dz}]_{k_d}$, the name of D and the public constructor $(P_z, Q_z)$, then he forwards them to D inside a Create D command. \item [8.] D derives $k_{dz}$ and responds with $[k_{dz}]_{k_z}$ to C as a Created D command. Then C encrypts it with $k_{cy}$ and forwards it to B as a Relay C command, then C forwards it to A as a Relay B command.
    \item [9.] Alice decrypts the packet and derives $k_{dz}$, the secret key between A and D.
\end{itemize}

\begin{figure}[htbp]
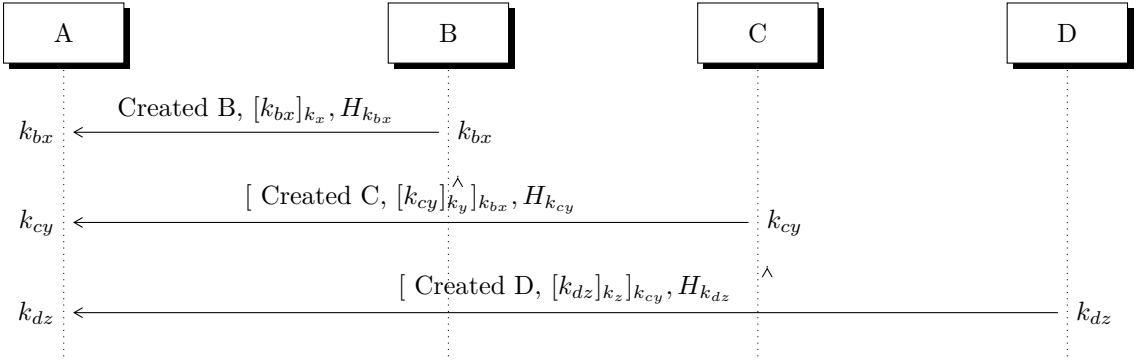
 
\centering
\begin{sequencediagram}
\newinst{A}{A}
\newinst[3.5]{B}{B}
\newinst[2.5]{C}{C}
\newinst[2.5]{D}{D}
\bloodymess[0]{B}{Created B, $[k_{bx}]_{k_x}, H_{k_{bx}}$}{A}{L}{$k_{bx}$}{$k_{bx}$}
\bloodymess[0.5]{B}{}{B}{L}{}{}
\bloodymess[0]{C}{$[$ Created C, $[k_{cy}]_{k_y}]_{k_{bx}},H_{k_{cy}}$}{A}{L}{$k_{cy}$}{$k_{cy}$}
\bloodymess[0.5]{C}{}{C}{L}{}{}
\bloodymess[0]{D}{$[$ Created D, $[k_{dz}]_{k_z}]_{k_{cy}}, H_{k_{dz}}$}{A}{L}{$k_{dz}$}{$k_{dz}$}
\end{sequencediagram}
\caption{Hash integrity check. The packets in the opposite direction contain the Hash value of the shared key.}
\label{fig:Fig.7}
\end{figure}

In the protocol has not been appended the hash value of the shared key into the packet response. We incorporates this functionality in the reverse packets depicted in Fig.~\ref{fig:Fig.7}.

\section{Discussion}
\label{Discussion}

After the third round of evaluation has been completed, NIST has chosen seven algorithms and eight alternative methods. Four of them are public key encryption and key-establishment systems. Three algorithms correspond to digital signature. In the first category are CRYSTALS-KYBER, NTRU-HPS, SABER which are lattice-based, while Classic McEliece is a public key encryption system based on code theory. In relation to digital signature schemes, CRYSTALS-DILITHIUM and FALCON are lattice-based and Rainbow is a multivariate-based method. Since our approach falls into the first category, we found that public keys size in Lizama's protocol has the smallest size: 0.256 kilobytes when $n$ reaches 1024 bits (see Table~\ref{tab:Table510}). 

Even more in~\cite{lizama2021non} has been demonstrated that Lizama's protocol can be scaled to support digital certificates and interoperability across different certification domains. For the reasons discussed above, we consider our approach to be very promising to achieve user anonymity in data networks in the pre-quantum and quantum era, since the size of the keys is highly competitive leading to greater computing speed and efficiency.

\begin{table}[htbp] 
\begin{center}
\caption{A comparison of Lizama's protocol in relation to the National Institute of Standards and Technology (NIST) Round 3 finalist algorithms is shown in the categories of public key encryption and key-establishment methods~\cite{NIST_Round_3}.}
\begin{tabular}{c l c c c } \\	
Scheme & System & Public Key (KB) & Private Key (KB) & Signature (KB)\\
\hline 

\multirow{4}{*}{\shortstack[l]{Public Key/ \\ KEM}} & \scriptsize{\scriptsize{LIZAMA'S KEP}} & $0.256-0.512$ & $0.192-0.384$ & -- \\

& \scriptsize{Classic McEliece} & $261,120 - 1,357,824$ & $6,492 - 14,120$ & -- \\

& \scriptsize{CRYSTALS-KYBER} & $1.632 - 3.168$ &  $0.8 - 1.568$ & -- \\
& \scriptsize{NTRU-HPS} & $0.931 - 1.230$ & $1.235 - 1.592$ & -- \\

& \scriptsize{SABER} & $0.672-1.312$ & $1.568-3.040$ & -- \\

& & & & \\

\multirow{3}{*}{\shortstack[l]{Signature \\ Algorithms}} & \scriptsize{CRYSTALS-DILITHIUM} & $1.312 - 2.592$ & -- & $2.420 - 4.595$\\
& \scriptsize{FALCON} & $0.897 - 1.793$ & -- & $0.666 - 0.280$\\
& Rainbow & $157.8 - 1,885.4$ & $101.2 - 1,375.7$ &  $0.066 - 0.212$ \\

\hline
\end{tabular}
\label{tab:Table510}
\end{center}
\end{table}

\begin{comment}
Lizama's ni-KEP & $\sim 0.2$ \\
Lizama's certified & $\sim 0.3$ \\
ECDSA & $\sim 0.1$ \\
RSA & $\sim 0.5$ \\
Code based & 190 \\
Lattice based & 11 \\
Multivariate & 99 \\
SS Isogenies & 122 \\ 
& & & & \\
\end{comment}

\section{Conclusions}
\label{Conclusions}

In this research we have discussed a new approach to circuit-based communication to achieve user anonymity through Lizama's non-invertible key exchange protocol. Since the circuit communication is the foundation of the onion routing protocol, it can be properly enhanced to anonymize web services as well. 

The non-invertible key exchange algorithm has been conceived on the basis of perfect secrecy, thus the approach presented here can be properly used in the pre-quantum and quantum era since Lizama's protocol has the smallest key size when compared to main post-quantum schemes. Furthermore Lizama's protocol can be scaled to support digital certificates and interoperability across different certification domains.

\appendix
\section{Lizama's Key Exchange Protocol}
\label{Lizama's KEP}

\begin{figure}[htbp]
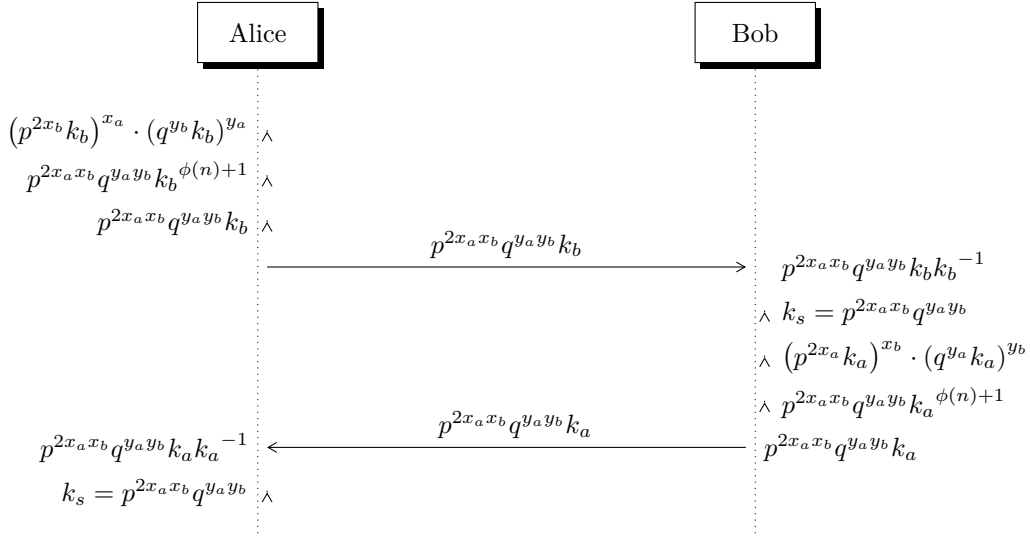
 
\centering
\begin{sequencediagram}
\newinst{A}{Alice}
\newinst[5]{B}{Bob}
%\bloodymess[1]{A}{$2^{2x_a} k_a$, $2^{(2r-1-x_a)} k_a$}{B}{R}{}{}
\bloodymess[0]{A}{}{A}{L}{}{${\left(p^{2x_b} k_b \right)}^{x_a}\cdot{\left(q^{y_b} k_b\right)}^{y_a}$}
\bloodymess[0]{A}{}{A}{R}{$p^{2 x_a x_b} q^{y_a y_b} {k_b}^{\phi(n)+1}$}{}
\bloodymess[0]{A}{}{A}{R}{$p^{2 x_a x_b} q^{y_a y_b} {k_b}$}{}
\bloodymess[0]{A}{$p^{2 x_a x_b} q^{y_a y_b} k_b$}{B}{R}{}{$ $ $ $ $p^{2 x_a x_b} q^{y_a y_b} k_b {k_b}^{-1}$}
\bloodymess[0]{B}{}{B}{L}{$ $ $ $ $k_s = p^{2 x_a x_b} q^{y_a y_b}$}{}
\bloodymess[0]{B}{}{B}{L}{$ $ $ $ ${\left(p^{2x_a} k_a\right)}^{x_b}\cdot{\left(q^{y_a} k_a\right)}^{y_b}$}{}
\bloodymess[0]{B}{}{B}{L}{$ $ $ $ $p^{2x_a x_b} q^{y_a y_b} {k_a}^{\phi(n)+1}$}{}
\bloodymess[0]{B}{$ $ $ $ $p^{2x_a x_b} q^{y_a y_b} k_a$}{A}{L}{$p^{2x_a x_b} q^{y_a y_b} {k_a}$}{$p^{2x_a x_b} q^{y_a y_b} k_a {k_a}^{-1}$}
\bloodymess[0]{A}{}{A}{R}{$k_s = p^{2x_a x_b} q^{y_a y_b}$}{}
\end{sequencediagram}
\caption{Lizama's non-invertible KEP~\cite{lizamanon}. All operations are modulo $n$ where $n=pqr$. According to Euler's theorem ${k}^{\phi(n)+1} \ \text{mod} \ n = k$ because $k$ is an invertible integer in $\mathbb{Z}_{n}$.}
\label{fig:Fig.30}
\end{figure}

Lizama's key exchange protocol was introduced in~\cite{lizamanon,lizama2021non} and is illustrated in Fig.~\ref{fig:Fig.30}. The public key of user $i$ ($a$ for Alice, $b$ for Bob) has two components $(P_i, Q_i)$ where $P_i = p^{2x_i} k_i \ \text{mod} \ n$ and $Q_i = q^{y_i} k_i \ \text{mod} \ n$. The value $x_i$ is chosen randomly while $y_i$ is computed according to the relation $y_i=\phi(n)-x_i+1$. The module $n$ is the product of three public integer primes, so that $n= p \cdot q \cdot r$ where $p$ and $q$ are small prime numbers and $r$ is a big integer prime. To achieve indistinguishability $p$ and $q$ are suggested to be 2, since 2 is a primitive root module $r$~\cite{lizamanon}. The exponent is chosen to be $2x_i$ instead of $x_i$ to avoid a multiplication attack. The $x_i$ value constitutes along $k_i$ the private key of user $i$ where $k_i$ is an invertible integer in the ring $\mathbb{Z}_{n}$. Users exchange their public keys $(P_i, Q_i)$ as well as the integer module $n$. The steps of the protocols are summarized as follows:

\begin{itemize}
\item [1.] Once public keys have been exchanged, the users perform two operations over the numbers received: exponentiation and multiplication as indicated in Tab.~\ref{tab:Table10}. 

\begin{table}[htbp] 	
\begin{center}
\caption{Exponentiation and multiplication are performed by users after their public keys have been exchanged.}
\begin{tabular}{|l|c|c|}
\hline
User 		& Operation  & Result \\  \hline
\multirow{2}{*}{Alice \ }   	& \multirow{2}{*}{$ \ \ {\left( {p}^{2x_b} \cdot k_b \ \text{mod} \ n \right)}^{x_a} \cdot {\left(q^{y_b} \cdot k_b\ \text{mod} \ n \right)}^{y_a} = $ \hspace{2mm}} & \multirow{2}{*}{\hspace{2mm} $p^{2 x_b x_a} q^{y_b y_a} \cdot k_b \ \text{mod} \ n $ \hspace{2mm}}\\ 
& & \\
\multirow{2}{*}{Bob}   	& \multirow{2}{*}{$ \ \ {\left( p^{2x_a} \cdot k_a \ \text{mod} \ n \right)}^{x_b} \cdot {\left(q^{y_a} \cdot k_a\ \text{mod} \ n \right)}^{y_b} = $ \hspace{2mm}} & \multirow{2}{*}{\hspace{2mm} $p^{2 x_a x_b} q^{y_a y_b} \cdot k_a \ \text{mod} \ n $ \hspace{2mm}}\\ 
& & \\
\hline
\end{tabular}
\label{tab:Table10}
\end{center}
\end{table}

\item [2.] To derive the results in the right column of Table~\ref{tab:Table10}, Euler's theorem is applied in $\mathbb{Z}_{n}$. The theorem is written in Eq.\ref{eq:Eq.1} where $n=pqr$ and $\phi(n)= (p-1)(q-1)(r-1)$. Here, $k$ and $n$ are relative prime each other, so $k$ is an invertible integer in $\mathbb{Z}_{n}$. Thus, according to Eq.\ref{eq:Eq.1} we have ${k}^{\phi(n)+1} = {k}^{\phi(n)} \cdot {k}^{1} = k$. 

\begin{equation}
\label{eq:Eq.1}
 {k}^{\phi(n)} \equiv 1 \ \text{mod} \ n
\end{equation}

\item [3.] Users exchange the resulting value $p^{2 x_a x_b} q^{y_a y_b} k_i \ \text{mod} \ n$, which is multiplied by the corresponding inverse ${k_i}^{-1}$ at each side to derive the secret shared key $p^{2x_a x_b} q^{y_a y_b} \allowbreak \ \text{mod} \ n$ as depicted in Fig.~\ref{fig:Fig.30}. 
\end{itemize}

%Let us write out an example about the required keys size from ~\cite{lizamanon,e23020226}. Consider that $p=q=2$ and $|r|=1024$, the length of the private key yields 1536 bits ($|x|=512$ and $|k|=1024$) while the public key $(P_i, Q_i)$ contains 2056 bits. The security level of the secret key is 1024. The process to determine the size of the key is the following: $P_i = p^{2x} \cdot k \ \text{mod} \ n$ thus $P_i = p^{2x} \ \text{mod} \ n \cdot k \ \text{mod} \ n$, which in turn implies that $|k|=|n|$. If $p=2$ and $n=4r$, we have $2^{2x} \ \text{mod} \ 4r$, then $4^{x} \ \text{mod} \ 4r$ yields $|4| \cdot |x| = |4| + |r|$ and $|x| \sim \frac{|r|}{2}$. Since the private key is conformed by $x$ and $k$, its size is computed as $|n|+|x| \sim |r|+|x|$ which gives 1536.

\subsection{Encryption-system}

In Fig.~\ref{fig:Fig.3}, the secret shared key $k_s$ is a non-invertible number in $\mathbb{Z}_n$, thus a convenient method to achieve a cipher-system and secret communication is to divide $k_s = p^{2x_a x_b} q^{y_a y_b} \ \text{mod} \ n$ by $pq$. Now, Alice and Bob can compute its multiplicative inverse ${k_r}^{-1}$. The enciphered message is obtained as $c = m \cdot k_r \ \text{mod} \ r$ and the original plaintext is recovered as $m = c \cdot {k_r}^{-1} \ \text{mod} \ r$ because $m = m \cdot k_r {k_r}^{-1} \ \text{mod} \ r$. To send a message encoded as an integer in $\mathbb{Z}_{r}$, the number $m$ must be less than $r$.

\begin{table}[htbp] 
\begin{center}
\caption{Encryption/decryption mathematical relations. 
}
\begin{tabular}{l   c}
\\	
Message & \hspace{3mm} Mathematical relation \hspace{3mm} \\
\hline 
Encryption  &  $c = m \cdot k_{r} \ \ \text{mod} \ \ r$ \\ \\
Decryption  &  $m = c \cdot {k_{r}}^{-1} \ \ \text{mod} \ \ r$ \\ 
\hline
\end{tabular}
\label{tab:Table1}
\end{center}
\end{table}

%\acknowledgements
%\label{sec:ack}
%At the end of the manuscript, right before the bibliography you might want to place an acknowledgment. This can be easily done by using the command \verb!\acknowledgements! as you can see here.

%\nocite{*}
%\bibliographystyle{abbrvnat}
% use the following instead if you encounter problems 
%\bibliographystyle{alpha}
\bibliographystyle{unsrt}

\bibliography{TOR_KEP}
\label{sec:biblio}

\end{document}